\documentclass[pra,aps,twocolumn,superscriptaddress,groupedaddress]{revtex4}
\usepackage{psfig}

\begin{document}

\title{Dissipation in systems of linear and nonlinear quantum scissors}

\author{Adam Miranowicz}

\address{Institute of Physics, Adam Mickiewicz University,
ul. Umultowska 85, 61-614 Pozna\'n, Poland}

\address{CREST Research Team for Interacting Carrier
Electronics, Graduate University for Advanced Studies (SOKENDAI),
Hayama, Kanagawa 240-0193, Japan}

\author{Wies\a law Leo\'nski}

\address{Institute of Physics, Adam Mickiewicz University,
ul. Umultowska 85, 61-614 Pozna\'n, Poland}

\date{Nov 2003, {\sc to appear in J. Optics B in March/April 2004}}

\begin{abstract}
We analyze truncation of coherent states up to a single-photon
Fock state by applying linear quantum scissors, utilizing the
projection synthesis in a linear optical system, and nonlinear
quantum scissors, implemented by periodically driven cavity with a
Kerr medium. Dissipation effects on optical truncation are studied
in the Langevin and master equation approaches. Formulas for the
fidelity of lossy quantum scissors are found.

\vspace{3mm} \noindent {\bf Keywords:} Quantum state engineering,
Kerr effect, qubit generation, finite-dimensional coherent states

% \submitto{\JOB}

\end{abstract}

\maketitle

%\vspace*{4mm}

\section{Introduction}

The breathtaking advances in quantum computation and quantum
information processing in the last decade \cite{Nie00} have
stimulated progress in quantum optical state generation and
engineering \cite{JMO97}. Among various schemes, the proposal of
Pegg, Phillips and Barnett \cite{Peg98,Bar99} of the optical state
truncation via projection synthesis has attracted considerable
interest
%\cite{Bab02,Res02,Vil99,Par00,Ozd01,Ozd02a,Ozd02b,Kon00,Vil01,Leo94,Dar00,Leo97,Mir96,Mir01,Leo01,Mir03}
\cite{Bab02}--\cite{Mir03} due to the simplicity of the scheme to
generate and teleport `flying' qubits defined as a running wave
superpositions of zero- and single-photon states. The scheme is
referred to as {\em linear quantum scissors} (LQS) since the
coherent state entering the system is truncated in its Fock
expansion to the first two terms using only linear optical elements
and performing conditional photon counting. The optical state
truncation can also be realized in systems comprising nonlinear
elements including a Kerr medium \cite{Leo94,Dar00}. Such systems
will be referred to as {\em nonlinear quantum scissors} (NQS). Both
LQS \cite{Kon00,Vil01} and NQS \cite{Leo97,Mir96} can be
generalized for the generation of a superposition of $N$ states. It
is worth noting that there are fundamental differences between the
states truncated by the LQS and NQS \cite{Mir01}.

In this paper we analyze the effects of dissipation on state
truncation by quantum scissors. Various kinds of losses in quantum
scissors have already been analyzed including: inefficiency and
dark counts of photodetectors
\cite{Bar99,Vil99,Par00,Ozd01,Ozd02a}, non-ideal single-photon
sources \cite{Ozd01,Ozd02a},  mode mismatch \cite{Ozd02b}, and
losses in beam splitters \cite{Vil99}. \"Ozdemir {\em et al.}
\cite{Ozd01,Ozd02a,Ozd02b} demonstrated that the LQS exhibit
surprisingly high fidelity in realistic setups even with
conventional photon counters so long as the amplitude of the input
coherent state is sufficiently small. LQS has recently been
realized experimentally by Babichev {\em et al.} \cite{Bab02} and
Resch {\em et al.} \cite{Res02}, although only in the low-intensity
regime. The effect of losses on the optical truncation in NQS has
been studied for zero-temperature reservoir \cite{Leo01} and
imperfect photodetection \cite{Dar00}.

These studies of losses in LQS (with few exceptions, e.g., for
\cite{Vil99}) have been based on the quantum detection and
estimation theory using the positive operator valued measures
(POVM) \cite{Hel76}. In quantum-optics textbooks (see, e.g.,
\cite{Lou73,Per91}), the quantum-statistical properties of
dissipative systems are usually treated in three ways, by applying
(i) the Langevin (Langevin-Heisenberg) equations of motion with
stochastic forces, (ii) the master equation for the density matrix,
and (iii) the classical Fokker-Planck equation for quasiprobability
distribution. In the next section we will apply the Langevin
approach to describe dissipative LQS, while in section 3 we shall
use the master equation approach to study dissipative NQS.

\section{Lossy linear quantum scissors in the Langevin approach}

The linear quantum scissors device of Pegg, Phillips and Barnett
\cite{Peg98,Bar99} is a simple physical system for optical state
truncation based only on {\em linear} optical elements (two beam
splitters BS1 and BS2) and two photodetectors (D2 and D3) as
depicted in figure 1. If the input modes $a_{1}$ and $a_{2}$ are in
the single-photon and vacuum states, respectively, and one photon
is detected at D2 but no photons at D3, then the lossless LQS
device with 50/50 beam splitters truncates the input coherent state
$|\alpha \rangle$ in mode $b_{3}$ to the following superposition of
vacuum and single-photon states in mode $b_{1}$
\begin{equation}
|\alpha _{{\rm trunc}}{\rangle }_{b1}=N^{\prime }\ _{c2c3}\langle
10|\psi _{ {\rm out}}{\rangle }_{b1c2c3}=\frac{|0\rangle
_{b1}+\alpha |1\rangle _{b1}}{ \sqrt{1+|\alpha |^{2}}} \label{M1}
\end{equation}
where $\alpha $ is the complex amplitude and $N^{\prime }$ is a
renormalization constant. The state (\ref{M1}) is referred to as
the truncated two-dimensional (or two-level) coherent state since
it is the normalized superposition of the first two terms of Fock
expansion of the Glauber coherent state. By introducing a new
variable $\bar{\alpha}$ such that $\cos(|\bar{\alpha}|)=1/
\sqrt{1+|\alpha |^{2}}$ and $\sin(|\bar{\alpha}|)=|\alpha |/
\sqrt{1+|\alpha |^{2}}$, and $\varphi={\rm Arg}\alpha$, state
(\ref{M1}) can be rewritten as
\begin{equation}
|{\alpha}_{\rm trunc}\rangle = \cos (|\bar{\alpha}|)|0\rangle
+e^{i\varphi} \sin (|\bar{\alpha}|)|1\rangle \label{M1a}
\end{equation}
where, for brevity, subscript $b_1$ is skipped. If the $j$th
($j=1,2$) beam splitter has an arbitrary but real transmission
coefficient $t_{j}$ and an imaginary reflection coefficient
$r_{j}$, then the LQS generates the state \cite{Ozd01}
\begin{equation}
|\psi {\rangle }_{b1}=\frac{|r_{1}t_{2}||0\rangle _{b1}+\alpha
|r_{2}t_{1}||1\rangle _{b1}}{\sqrt{|r_{1}t_{2}|^{2}+|\alpha
|^{2}|r_{2}t_{1}|^{2}}} . \label{M2}
\end{equation}
This state evolves into the truncated coherent state (\ref{M1}) by
assuming identical BSs ($r_{1}=r_{2}$ and $t_{1}=t_{2}$).

In general, the transmission and reflection coefficients of a
perfect BSs obey the conditions $\left| t\right| ^{2}+\left|
r\right| ^{2}=1$ and $tr^{\ast }+t^{\ast }r=0$, implied by the
unitarity of BS transformation. By including dissipation, these
conditions can be violated. Thus, the main goal of this section is
to analyze the deterioration of the truncation process due to the
noise introduced by lossy beam splitters and also by inefficient
photodetectors. In the simplest approach, one can model the BS
losses and finite detector efficiency by adding to our system
additional beam splitters, then all components of the system
(including the new BSs) can be assumed perfect. Here, we apply
another standard approach of the quantum theory of damping based on
the Langevin noise operators \cite{Lou73,Per91}. We follow the
analyses of Barnett {\em et al.} \cite{Bar96,Bar98} and
Villas-B\^{o}as {\em et al.} \cite{Vil99}. The lossy BS1 transforms
the input annihilation operators $\hat{a}_{j}$ into the output
$\hat{b}_{j}$ as follows \cite{Bar96,Bar98}
\begin{eqnarray}
\hat{a}_{1} &=&t_{1}^{\ast }\hat{b}_{1}+r_{1}^{\ast }
\hat{b}_{2}+\hat{L}_{a1} ,\nonumber \\
\hat{a}_{2} &=&r_{1}^{\ast }\hat{b}_{1}+t_{1}^{\ast
}\hat{b}_{2}+\hat{L}_{a2} \label{M3}
\end{eqnarray}
where we use the notation of figure 1, and $\hat{L}_{a1}$ and
$\hat{L}_{a2}$ are the Langevin noise (force) operators satisfying
the following commutation relations
\begin{eqnarray}
\lbrack \hat{L}_{a1},\hat{L}_{a1}^{\dag }]
&=&[\hat{L}_{a2},\hat{L}_{a2}^{\dag
}]=1-|t_{1}|^{2}-|r_{1}|^{2}\equiv \Gamma _{1} ,\nonumber \\
\lbrack \hat{L}_{a1},\hat{L}_{a2}^{\dag }]
&=&[\hat{L}_{a2},\hat{L}_{a1}^{\dag }]=-(t_{1}r_{1}^{\ast
}+t_{1}^{\ast }r_{1})\equiv -\Omega _{1} . \label{M4}
\end{eqnarray}
The transformation between the input ($\hat{b}_{j}$) and output
($\hat{c}_{j}$) annihilation operators of the lossy BS2 together
with effect of finite efficiency ($\eta \equiv \eta _{1}=\eta
_{2}$) of detectors generalizes to \cite{Vil99}
\begin{eqnarray}
\hat{b}_{2} &=&\sqrt{\eta }t_{2}^{\ast }\hat{c}_{2}+\sqrt{\eta
}r_{2} ^{\ast }\hat{c}_{3}+\hat{L}_{b2},
\nonumber \\
\hat{b}_{3} &=&\sqrt{\eta }r_{2}^{\ast }\hat{c}_{2}+\sqrt{\eta
}t_{2}^{\ast }\hat{c}_{3}+\hat{L}_{b3} \label{M5}
\end{eqnarray}
where the Langevin noise operators $\hat{L}_{b2}$ and
$\hat{L}_{b3}$ obey
\begin{eqnarray}
\lbrack \hat{L}_{b2},\hat{L}_{b2}^{\dag }]
&=&[\hat{L}_{b3},\hat{L}_{b3}^{\dag }]=\eta \Gamma
_{2}+(1-\eta )\equiv x ,\nonumber \\
\lbrack \hat{L}_{b2},\hat{L}_{b3}^{\dag }]
&=&[\hat{L}_{b3},\hat{L}_{b2}^{\dag }]=-\eta \Omega _{2} .
\label{M6}
\end{eqnarray}
In (\ref{M4}) and (\ref{M6}), $\Omega _{j}=t_{j}r_{j}^{\ast
}+t_{j}^{\ast }r_{j}$ and $ \Gamma _{j}=1-|t_{j}|^{2}-|r_{j}|^{2}$
are the $j$th beam splitter phase and amplitude dissipation
coefficients, respectively, which vanish for perfect beam
splitters. For simplicity, we assume that the BSs are identical ($
r_{1}=r_{2}\equiv r,$ $t_{1}=t_{2}\equiv t$) and they cause only
amplitude damping ($\Gamma \equiv \Gamma _{1}=\Gamma _{2}\neq 0)$
without introducing phase noise ($\Omega _{1}=\Omega _{2}=0$). By
applying the transformations (\ref{M3}) and (\ref{M5}) for the
input state $|\psi _{{\rm in}}{\rangle } _{a1a2b3}=|1\rangle
_{a1}|0{\rangle }_{a2}|\alpha {\rangle }_{b3}$ and performing the
conditional measurement (projection synthesis) on modes $c_{2}$ and
$c_{3}$ (as shown in figure 1), one finds that the state of the
output mode $b_{1}$ of the LQS is entangled with the environment as
follows \cite{Vil99}
\begin{eqnarray}
|\psi {\rangle }_{b1E}&=&N^{\prime \prime }\ _{c2c3}\langle
10|\psi _{{\rm
out}}{\rangle }_{b1c2c3E}  \nonumber \\
&=& N(|0{\rangle }_{b1}|\mathbf{\Lambda }_{0}\rangle _{E}+\alpha
|1{\rangle } _{b1}|\mathbf{\Lambda }_{1}\rangle _{E}) \label{M7}
\end{eqnarray}
where we write compactly the environmental states as
\begin{eqnarray}
|\mathbf{\Lambda }_{0}\rangle _{E} &=&\sqrt{\eta }r(t+\alpha
r\hat{L}_{a2}^{\dag }+\alpha \hat{L}_{a1}^{\dag })\exp (\alpha
\hat{L}_{b3}^{\dag })|\mathbf{0}{\rangle }_{E} ,
\nonumber \\
|\mathbf{\Lambda }_{1}\rangle _{E} &=&\sqrt{\eta }rt\exp (\alpha
\hat{L}_{b3}^{\dag })|\mathbf{0}{\rangle }_{E} \label{M8}
\end{eqnarray}
and the normalization $N$ is given by
\begin{equation}
N=\{\eta |r|^{2}\ |\alpha |^{2}e^{x|\alpha |^{2}}[|t|^{2}(|\alpha
|^{-2}+1)+|r|^{2}x+\Gamma ]\}^{-1/2} \label{M9}
\end{equation}
and $N^{\prime \prime }$ is a renormalization constant. The
fidelity of the output state (\ref{M7}) of the lossy LQS to a
desired perfectly truncated state, given by (\ref{M1}), can be
calculated from
\begin{equation}
F\equiv ||_{b1}\langle \alpha _{{\rm trunc}}|\psi {\rangle
}_{b1E}||^{2} \label{M10}
\end{equation}
which leads us to the following relation
\begin{eqnarray}
F &=&N^{2}\eta |r|^{2}\ \exp (x|\alpha |^{2})  \nonumber \\
&&\times \left( |t|^{2}(|\alpha |^{2}+1)+\frac{|\alpha
|^{2}}{1+|\alpha |^{2} }(|r|^{2}x+\Gamma )\right) \label{M11}
\end{eqnarray}
where the normalization $N$ is given by (\ref{M9}). By defining
$R=1/|\alpha |^{2}$, equation (\ref{M11}) can be simplified to
\begin{equation}
F=1-\frac{(\eta \Gamma +1-\eta )|r|^{2}+\Gamma }{(1+R)[(\eta
\Gamma +1-\eta )|r|^{2}+\Gamma +|t|^{2}(1+R)]} \label{M12}
\end{equation}
where $x=\eta \Gamma +(1-\eta )$. In a special case for 50/50 BSs,
$ |r|^{2}=|t|^{2}$, our solution simplifies to that of
Villas-B\^{o}as {\em et al.} (\cite{private}, note that the
corresponding fidelity in \cite{Vil99} is misprinted). By
neglecting losses caused by beam splitters, solution (\ref{M12})
is further reduced to the well-known Pegg-Phillips-Barnett
fidelity \cite{Peg98}
\begin{equation}
F=1-\frac{|\alpha |^{4}(1-\eta )}{(1+|\alpha |^{2})[1+|\alpha
|^{2} (2-\eta )] }. \label{M13}
\end{equation}
By assuming also perfect detectors, the fidelity becomes unity, as
expected.

\begin{figure}
\hspace*{5mm}\psfig{figure=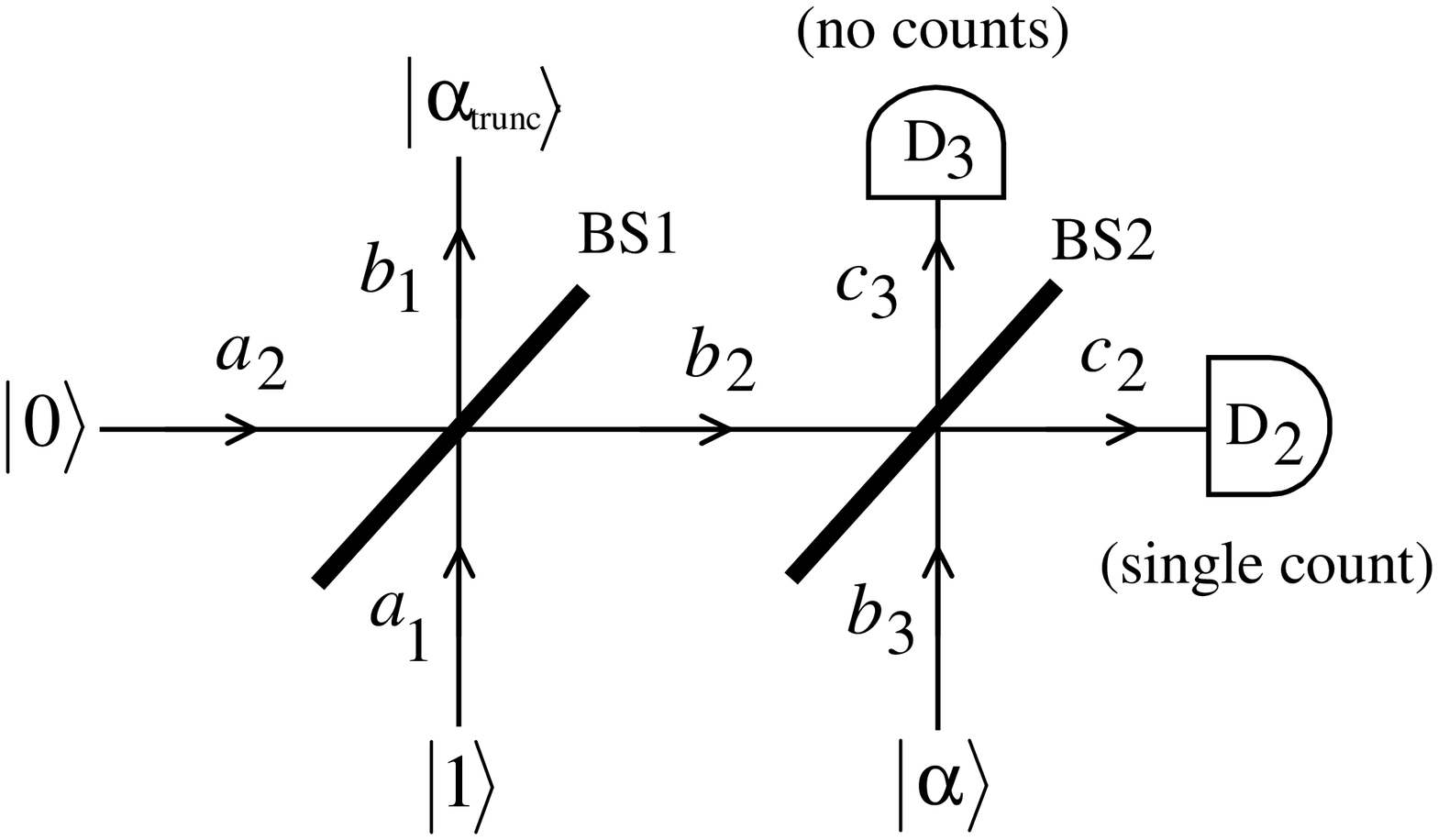,width=9cm}\vspace*{-2.5cm}
 \caption{Scheme of linear quantum scissors:
$|\alpha\rangle$ is the input coherent state, $|\alpha_{\rm
trunc}\rangle$ is the output truncated coherent state, $|0\rangle$
and $|1\rangle$ are vacuum and single-photon states, respectively;
BS1 and BS2 are beam splitters; D2 and D3 are photon detectors.
For successful truncation process, one of the detectors should
detect one photon, while the other -- no photons.}
\end{figure}
\begin{figure}
\vspace*{-5mm}\hspace*{15mm}\psfig{figure=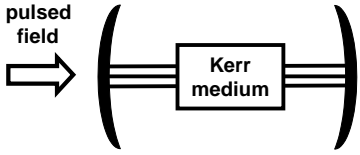,width=9cm}\vspace*{-4cm}
\caption{Scheme of nonlinear quantum scissors: cavity with a Kerr
medium is pumped by external ultra-short pulses of laser light. }
\end{figure}

\section{Lossy nonlinear quantum scissors in the master equation approach}

In nonlinear quantum scissors scheme, schematically depicted in
figure 2, a cavity mode is pumped by an external classical pulsed
laser field, described by Hamiltonian $\hat{H}_{K}$, and is
interacting with a Kerr medium, described by Hamiltonian
$\hat{H}_{NL}$. Thus, the whole system Hamiltonian is given by
\cite{Leo94,Leo01}
\begin{equation}
\hat{H}_{S}=\hat{H}_{0}+\hat{H}_{NL}+\hat{H}_{K}  \label{N01}
\end{equation}
where
\begin{eqnarray}
\hat{H}_{NL} &=&\hbar \frac{\kappa }{2}(\hat{a}^{\dag })^{2}\hat{a}^{2}
, \label{N01a} \\
\hat{H}_{K} &=&\hbar \epsilon (\hat{a}^{\dag
}+\hat{a})\sum_{k=-\infty }^{\infty }\delta (t-kT_{K})
\label{N01b}
\end{eqnarray}
and the free Hamiltonian of the system is $\hat{H}_{0}=\hbar
\omega \hat{a}^{\dag }\hat{a}$. In Eq. (\ref{N01a}), $\hat{a}$ is
the annihilation operator for a cavity mode at frequency $\omega
$, and $\kappa $ is the nonlinear coupling proportional to the
third-order susceptibility of the Kerr medium. In (\ref{N01b}),
Dirac $\delta$ functions describe external ultra-short light
pulses (kicks); real parameter $\epsilon$ is the strength of the
interaction of the cavity mode with the external field; $T_{K}$ is
the period of free evolution between the kicks. The truncation
process in the system, given by (\ref{N01}), occurs if (i)
$T_{K}\gg T_{ \mathrm{round-trip}}\gg 2\pi /\omega $, where
$\omega $ is the light frequency and $T_{\mathrm{round-trip}}$ is
the round-trip time of the light in the cavity, and (ii) the kicks
are much weaker than the Kerr nonlinear interaction, $\epsilon \ll
\kappa $. As shown in Refs. \cite{Leo94,Leo97}, the state
generated by the NQS is a two-dimensional coherent state
\cite{Buz92,Mir94} of the form
\begin{equation}
|\bar{\alpha}_{\rm trunc}\rangle\approx \cos
(|\bar{\alpha}|)|0\rangle -i\sin (|\bar{\alpha}|)|1\rangle
\label{N01c}
\end{equation}
where $\bar{\alpha}=-ik\epsilon $. Dissipation of the NQS system
is modelled by its coupling to a reservoir of oscillators (heat
bath) described by the Hamiltonian
\begin{eqnarray}
\hat{H} &=&\hat{H}_{S}+\hat{H}_{R}+\hat{H}_{SR} ,  \label{N02a}
\\
\hat{H}_{SR} &=&\hbar \sum_{j}(g_{j}\hat{a}\hat{b}_{j}^{\dag
}+g_{j}^{\ast }\hat{a}^{\dag }\hat{b}_{j}) \label{N02b}
\end{eqnarray}
where $\hat{H}_{S}$ is given by (\ref{N01}) and $\hat{H}_{R}=\hbar
\sum_{j}\chi_{j}\hat{b}_{j}^{\dag }\hat{b}_{j}$ is the free
Hamiltonian of the reservoir, where $\hat{b}_{j}$ is the boson
annihilation operator of the $j$th reservoir oscillator. By
applying the standard methods of the quantum theory of damping
\cite{Lou73}, one finds that the NQS evolution between the kicks is
governed under the Markov approximation by the following master
equation in the interaction picture
\begin{equation}
\frac{\partial }{\partial t}\hat{\rho} =-i\frac{\kappa
}{2}[(\hat{a}^{\dag })^{2}\hat{a}^{2},\hat{\rho} ]-\frac{\gamma
}{2}([\hat{a}^{\dag },\hat{a}\hat{\rho} ]+{\rm h.c.})+\gamma
\bar{n}[\hat{a}^{\dag },[\hat{\rho} ,\hat{a}]] \label{N03}
\end{equation}
where $\gamma $ is the damping constant and $\bar{n}$ is the mean
number of thermal photons, $\bar{n}=\{\exp [\hbar \omega
/(k_{B}T)]-1\}^{-1}$, at the reservoir temperature $T$, where
$k_{B}$ is the Boltzmann constant. Let the kick be applied at time
$t_{K}$, then the solution of Eq. (\ref{N03}) for any time $t$
after $t_{K}$ but before moment $t_{K}+T_{K}$ is the same as the
solution for the ordinary damped anharmonic oscillator \cite
{Dan89,Per90,Gan91} with the initial state given at time $t_{K}$.
We can write the solution compactly as $(\rho_{nm}\equiv \langle
n|\hat{\rho}|m\rangle)$:
\begin{eqnarray}
\rho _{nm}(t_{K}+t) &=&\exp [\frac{\gamma t}{2}+i(n-m)\kappa
t]E_{n-m}^{n+m+1}(t)  \label{N04} \\
&&\times \sum_{l=0}^{\infty }\rho _{n+l,m+l}(t_{K})\sqrt{
C_{n}^{n+l}C_{m}^{m+l}}\bar{g}_{n-m}^{l}(t)  \nonumber \\
&&\times F[-n,-m,l+1;\frac{4\bar{n}(\bar{n}+1)}{\Delta
_{n-m}^{2}}\sinh ^{2}t_{n-m}]  \nonumber
\end{eqnarray}
where $F$ is the hypergeometric function, $C_{y}^{x}$ are binomial
coefficients, $t_{x}=\gamma \Delta _{x}t/2$ and
\begin{eqnarray}
\bar{g}_{x}(t) &=&\frac{2(\bar{n}+1)}{\Omega _{x}+\Delta _{x}\coth
t_{x}},
\nonumber \\
E_{x}(t) &=&\frac{\Delta _{x}}{\Omega _{x}\sinh t_{x}+\Delta _{x}\cosh t_{x}}
\label{N05}
\end{eqnarray}
with $\Delta _{x}=\sqrt{\Omega _{x}^{2}-4\bar{n}(\bar{n}+1)}$ and
$\Omega _{x}=1+2\bar{n}+i\kappa x/\gamma .$ By assuming the
reservoir to be at zero temperature, the solution (\ref{N04})
reduces to \cite{Mil86,Per88,Mil91}
\begin{eqnarray}
&&\rho _{nm}(\tau _{K}+\tau ) =\exp [i(n-m)\frac{\tau }{2}
]f_{n-m}^{(n+m)/2}(\tau )
\label{N06} \\
&&\quad \times \sum_{l=0}^{\infty }\rho _{n+l,m+l}(\tau _{K})
\sqrt{C_{n}^{n+l}C_{m}^{m+l}}\left( \frac{\lambda \lbrack
1-f_{n-m}(\tau )]}{\lambda +i(n-m)}\right) ^{l} \nonumber
\end{eqnarray}
where $\tau $ is the scaled time given by $\tau =\kappa t$, so
$\tau _{K}=\kappa t_{K}$. Moreover, $\lambda =\gamma /\kappa $,
and $f_{x}(\tau )=\exp [-(\lambda +ix)\tau ]$. For a lossless
anharmonic oscillator, i.e. for $ \lambda =0$, the solution
(\ref{N06}) further simplifies to
\begin{equation}
\rho _{nm}(\tau _{K}+\tau )=\exp \{i[n(n-1)-m(m-1)]\frac{\tau }{2}\}\rho
_{nm}(\tau _{K})  \label{N07}
\end{equation}
Solution (\ref{N04}) describes the evolution of the NQS between
the kicks only. On the other hand, the evolution at each kick is
given by
\begin{eqnarray}
&&\lim_{\delta \rightarrow 0}\langle n|\hat{\rho} (t_{K}+\delta
)|m\rangle
\label{N11} \\
&&\quad \quad =\lim_{\delta \rightarrow 0}\sum_{n^{\prime
},m^{\prime }=0}^{\infty }\!\!U_{nn^{\prime }}\langle n^{\prime
}|\hat{\rho} (t_{K}-\delta )|m^{\prime }\rangle U_{mm^{\prime
}}^{\ast }  \nonumber
\end{eqnarray}
where
\begin{equation}
U_{nm}=\langle n|\hat{U}|m\rangle =\langle n|\exp [-i\epsilon
(\hat{a}^{\dag }+\hat{a})]|m\rangle   \label{N12}
\end{equation}
in analogy to the Milburn-Holmes transformation for the pulsed
parametric amplifier with a Kerr nonlinearity \cite{Mil91}. By
observing that $\hat{U}$ is the displacement operator
$\hat{U}=\exp[-i\epsilon (\hat{a}^{\dag }+\hat{a})]=
\hat{D}(-i\epsilon )$, we can use the well-known Cahill-Glauber
\cite{Cah69} formulas leading for $n\geq m$ to
\begin{equation}
U_{nm}=e^{-\epsilon^{2}/2}\sqrt{\frac{m!}{n!}}(-i\epsilon
)^{n-m}L_{m}^{n-m}(\epsilon^{2})  \label{N13}
\end{equation}
and for $n<m$ to
\begin{equation}
U_{nm}=e^{-\epsilon^{2}/2}\sqrt{\frac{n!}{m!}}(i\epsilon
)^{m-n}L_{n}^{m-n}(\epsilon^{2})  \label{N14}
\end{equation}
where $L_x^{y}(z)$ is an associated Laguerre polynomial. Thus, we
have a complete solution to describe the effects of dissipation on,
in particular, the truncation fidelity after the $k$th kick, which
is given by
\begin{eqnarray}
\bar{F}(t)&=&\langle \bar{\alpha}_{\rm trunc}|\hat{\rho}
(t)|\bar{\alpha}_{\rm trunc}
\rangle  \label{N15} \\
&=&\cos ^{2}(k\epsilon)\rho _{00}(t)+\sin (2k\epsilon)\mathrm{Im}
\rho _{01}(t)+\sin ^{2}(k\epsilon)\rho _{11}(t)  \nonumber
\end{eqnarray}
where the perfectly truncated state $|\bar{ \alpha}\rangle
_{\mathrm{NQS}}$ was applied according to (\ref{N01c}).

\section{CONCLUSIONS}

We studied dissipative quantum scissors systems for truncation of
a Glauber (infinite-dimensional) coherent state to a superposition
of vacuum and single-photon Fock states (two-dimensional coherent
state). We have contrasted the Pegg-Phillips-Barnett quantum
scissors based on linear optical elements and the
Leo\'nski-Tana\'s quantum scissors comprising nonlinear Kerr
medium. We analyzed the effects of dissipation on truncation
fidelity in the linear scissors within the Langevin noise operator
approach and in the nonlinear system in the master equation
approach.

\noindent {\bf Acknowledgements}. AM warmly thanks \c{S}ahin K.
\"Ozdemir, Masato Koashi, and Nobuyuki Imoto for long and
stimulating collaboration on experimental realization of quantum
scissors. The authors also thank Ryszard Tana\'s for helpful
discussions.

% \section*{REFERENCES}

\end{document}